\documentstyle[epsfig]{mn} 

\title[Rotational Broadening and Doppler Tomography of Centaurus X-4]{Rotational Broadening and Doppler Tomography of the Quiescent X-Ray Nova Centaurus X-4}  
\author[M. A. P. Torres et al.]{M. A. P. Torres$^{1,2,3}$, J. Casares$^{2}$, I. G. Mart\'\i{}nez-Pais$^{1,2}$, P. A. Charles$^{4,5}$\\
$^{1}$Departamento de Astrof\'\i{}sica, Universidad de La Laguna, E-38271, La Laguna, Tenerife, Spain \\
$^{2}$Instituto de Astrof\'\i{}sica de Canarias, E-38200, La Laguna, Tenerife, Spain \\
$^{3}$Physics Department, University College, Cork, Ireland \\   
$^{4}$Department of Astrophysics, Nuclear Physics Laboratory, Keble Road, Oxford, OX1 3RH \\
$^{5}$Department of Physics and Astronomy, University of Southampton, Southampton, SO17 1BJ }                            

\begin{document}

\maketitle

\begin{abstract}

We present high and intermediate resolution spectroscopy of the X-ray nova Centaurus X-4 during its quiescent phase. Our analysis of the absorption features supports a ${\rm K}3-{\rm K}5$ V spectral classification for the companion star, which contributes $\simeq75$~per cent of the total flux at H$\alpha$. Using the high resolution spectra we have measured the secondary star's rotational broadening to be $v \sin i = 43 \pm 6$ km s$^{-1}$ and determined a binary mass ratio of $q=0.17 \pm 0.06$. Combining our results for $K_2$ and $q$ with the published limits for the binary inclination, we constrain the mass of the compact object and the secondary star to the ranges $0.49 < M_1 < 2.49$ M$_\odot$ and $0.04 < M_2 < 0.58$ M$_\odot$. A Doppler image of the H$\alpha$ line shows emission coming from the secondary star, but no hotspot is present. We discuss the possible origins of this emission.
 
\end{abstract}


\section{INTRODUCTION}

X-ray novae (XRNe) are a subclass of low mass X-ray binaries characterized by the exhibition of occasional outbursts with a recurrence time-scale of a few decades (see the reviews by van Paradijs \& McClintock 1995; Tanaka \& Shibazaki 1996; van Paradijs 1998). During the outburst state the optical spectrum of the companion star is overwhelmed by the light of the X-ray irradiated accretion disc surrounding the compact object. During the long intervals of quiescence, the X-ray luminosity falls by at least 5 orders of magnitude, rendering the spectrum of the secondary star detectable. This provides an ideal opportunity to obtain the system parameters, constrain the physical properties of both the compact object and companion star and hence cast light on the nature and origin of these binary systems (see e.g. Charles 1998). In this way it has been shown that $\sim~75$~per cent of the XRNe harbour black holes. 
  
Centaurus X-4 is a well-studied XRN. It was discovered during an X-ray outburst in 1969 (Conner, Evans \& Belian 1969), and a second outburst ten years later allowed the identification of its optical counterpart which had brightened by $\geq$~6 magnitudes (Canizares, McClintock \& Gridlady 1980). A Type I X-ray burst was detected during the decline phase of the outburst (Matsuoka et al. 1980), indicating that the primary in this case is an accreting neutron star. Photometric and spectroscopic studies of the optical counterpart in its quiescent state led to the determination of a 15.1 hr orbital period (Chevalier et al. 1989; McClintock \& Remillard 1990) and a $\sim$~0.2~M$_\odot$ mass function (Cowley et al. 1988). The secondary star was identified as a ${\rm K}5-{\rm K}7$ evolved main sequence star.

The Cen X-4 quiescent light curves show a double-humped modulation with a full amplitude of $\sim$~0.15~mag (Cowley et al. 1988; Chevalier et al. 1989; McClintock \& Remillard 1990) which is associated with the tidal distortion of the Roche-lobe-filling secondary star. McClintock \& Remillard (1990), based on the phasing of the H$\beta$ emission line radial velocity curve, suggested the presence of an accretion disc hotspot to explain the unequal maxima and unusual different depths of the minima in the {\it V} and {\it I} light curves. Star spots or X-ray heating of the K-star were also invoked to explain the long term changes observed in the light curves by Chevalier et al. (1989). Shahbaz, Naylor \& Charles (1993) observed Cen X-4 in the IR, where the contamination of the ellipsoidal light curve by the accretion disc is minimized. Their fit of the {\it H}-band light curve with an ellipsoidal model led to the determination of a low inclination ($31^{\circ}-54^{\circ}$) and a primary mass lying in the range $0.5-2.1$~M$_\odot$. 

In this paper we refine the system parameters and ephemeris of Cen X-4. From our higher resolution spectra we are able to determine the rotational broadening of the companion star for the first time. Finally, we investigate the H$\alpha$ Doppler map of the system during the quiescent state. 
 
\section{OBSERVATIONS AND DATA REDUCTION}

Centaurus X-4 in its quiescent state was observed during three runs. The first was from 1993 June 24-27 when spectra were acquired using the RGO spectrograph attached to the 3.9-m Anglo-Australian Telescope (AAT). A 1200 line mm$^{-1}$ grating centred on H$\alpha$ and a 1.2 arcsec slit width yielded a dispersion of 0.8 \AA~pix$^{-1}$ and a spectral resolution of 74 km s$^{-1}$ (FWHM). On 1994 March 6-8 we employed the EMMI spectrograph on ESO's New Technology Telescope (NTT) on La Silla. Use of grating \# 6 and a 1.2 arcsec slit yielded a dispersion of 0.31 \AA~pix$^{-1}$ and a spectral resolution of 45 km s$^{-1}$. Our final run was with the 4.2-m William Herschel Telescope (WHT) on La Palma. Spectra were acquired during the nights of 1994 July 9/10 and 13/14 using the red arm of ISIS, a 1200R grating and a 0.8 arcsec slit which provided a spectral resolution of 25 km s$^{-1}$. In addition to the spectra of Cen X-4, on the night of July 9 radial velocity standard stars of spectral types K3~III (HR 5536), K4~III (HR 6159) and K5~III (HR 5622) were acquired using the same setting as for the target. Table~\ref{tablelog} provides a more detailed journal of the Cen X-4 observations.

All the runs were photometric, with seeing less than 1.5 arcsec at all times, and $<1$ arcsec occasionally. Copper-Neon arc lamp spectra were taken every hour. The spectra were extracted in a standard way using {\sc iraf}\footnote{{\sc iraf} is distributed by the National Optical Astronomy Observatories.}. The pixel-to-wavelength calibration was derived from third-order polynomial fits to 12 arc lines, giving a rms scatter $\le0.03$~\AA. The calibration curves were interpolated in time and the accuracy of the whole process was checked by cross-correlating the sky spectra, which was found to be within 10 km s$^{-1}$. The spectra were re-binned into a constant velocity scale and normalized by dividing with the result of fitting a low order spline to the continuum after masking the emission lines.

\section{RADIAL VELOCITY CURVE AND ORBITAL EPHEMERIS}
   
\begin{figure}
\epsfig{width=6in,file=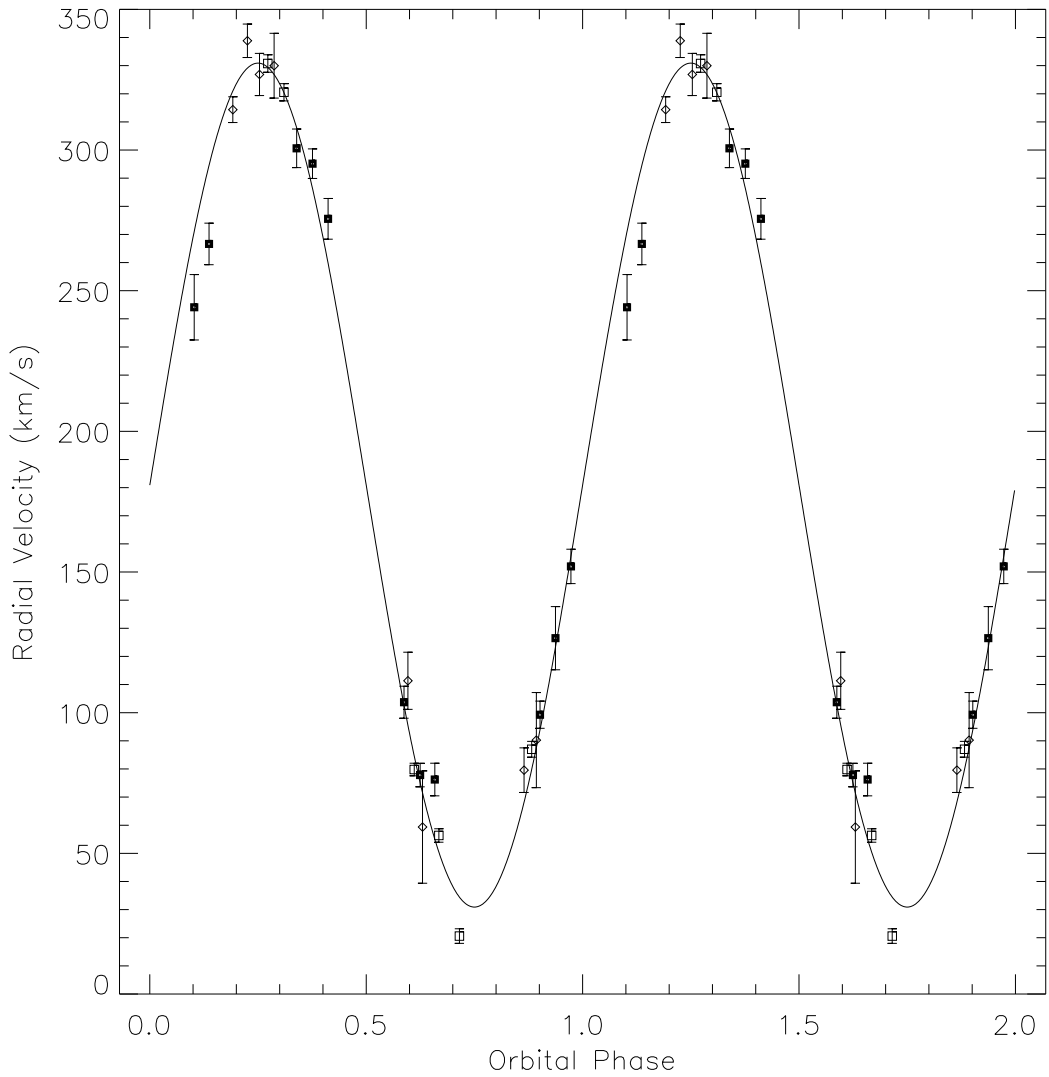} 
\caption{Radial velocities of the secondary star in Cen X-4 folded on the ephemeris of Sect. 3.1. Observations were performed in 1993 June (solid squares), 1994 March (squares) and 1994 July (diamonds). The best sine-wave fit is also shown.}
\label{fig1}
\end{figure}

Radial velocities were extracted by cross-correlating each target spectrum with the spectra of the three template stars in the range $\lambda\lambda6387-6513$. We performed least-squares sine fits to our radial velocity data, fixing the orbital period at $P=0.629063$ d (McClintock \& Remillard 1990). The resultant dynamical parameters are summarized in Table~\ref{tablerv}. We adopt the K4 III parameters of the radial velocity curve because it provided the lowest reduced $\chi^2$: $K_{2}$ = 150.0 $\pm$ 1.1 km s$^{-1}$, $\gamma_{_2}$ = 184.0 $\pm$ 1.9 km s$^{-1}$ and ${T_0}=\rm{HJD}~2449163.934\pm0.001$. This zero phase is defined as inferior conjunction of the secondary star. The combination of our $T_{\rm 0}$ determination with the value reported by McClintock \& Remillard (1990) enables us to refine the orbital period to $P=0.6290496\pm0.0000021$~d. Fig.~\ref{fig1} displays the radial velocity curve. Although our velocity semi-amplitude $K_{2}$ agrees with the values presented by Cowley et al. (1988) and McClintock \& Remillard (1990), we find a systemic velocity intermediate to these reported earlier.\footnote{These values are: $K_{2}$ = 152.0 $\pm$ 10 km s$^{-1}$, $\gamma$ = 234 $\pm$ 8 km s$^{-1}$ (Cowley et al. 1988). $K_{2}$ = 146 $\pm$ 12 km s$^{-1}$, $\gamma$ = 137 $\pm$ 17 km s$^{-1}$  (McClintock \& Remillard 1990).} This discrepancy could be due to the use of an erroneous heliocentric radial velocity for the template stars used by these authors. Unfortunately we have not found any updating of them.

\section{SPECTRAL CLASSIFICATION AND ROTATIONAL BROADENING}

We determined the spectral type and rotational broadening of the secondary by subtracting a set of templates from the Doppler-corrected average spectrum of the target until the lowest residual was obtained (see e.g. Marsh, Robinson \& Wood 1994).

Regarding the spectral classification we have constrained this analysis to these spectra acquired with the AAT because of their higher signal-to-noise ratio. The target spectra were summed in the rest frame of the template star HR 6159 by using the radial velocity parameters of Sect. 3. Different weights were assigned to different target spectra in order to maximize the signal-to-noise of the sum. In addition to the K~III-type standards acquired with the WHT, we have employed a set of K~V-type standards observed with the Isaac Newton Telescope (INT) in 1993 October (see Casares et al. 1996). These spectra were Doppler-shifted to the rest frame of HR 6159 and degraded appropriately to match the broadening of our AAT observations. This value was found by subtracting different broadened copies of the WHT and INT spectral templates from the Doppler-corrected average of Cen X-4 and performing a $\chi^2$ test on the residuals. 

\begin{figure}
\epsfig{width=6in,file=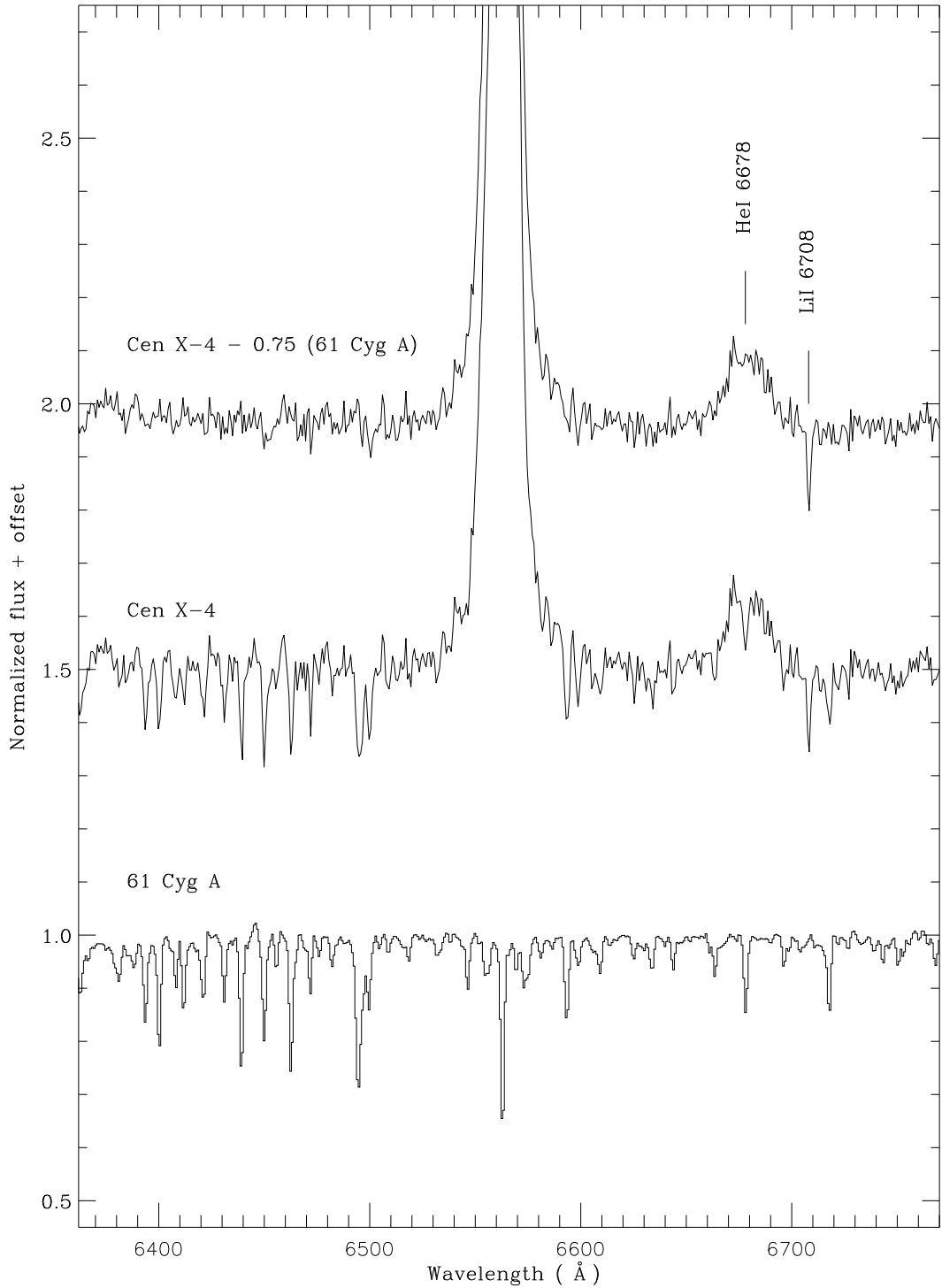}
\caption{Optimal subtraction of a K5~V template star (bottom spectrum) from the Doppler-corrected average of Cen X-4 in the rest frame of the secondary (middle spectrum). The residual of the subtraction (top spectrum) represents the relative contribution of the accretion disc. Arbitrary vertical offsets have
been applied to the spectra for the sake of clarity.}
\label{fig2}
\end{figure}

The templates were multiplied by a factor $0<f<1$, representing the fractional contribution of light from the secondary star, and subtracted from the target average. Then a $\chi^2$ test on the residuals was performed in the range $\lambda\lambda6387-6513$. The optimal values of $f$ are provided by minimizing $\chi^{2}$, with quoted uncertainties corresponding to $\chi^{2}_{\rm min}$+1 (Lampton, Margon \& Bowyer 1976). The minimization of $\chi^2_\nu$ does not give a single minimum (see Table~\ref{tablespeclas}) and the secondary star in Cen X-4 is most probably of spectral type ${\rm K}3-{\rm K}5$ contributing $\simeq75$~per cent to the total flux of the system at H$\alpha$. A similar result was obtained when applying this technique to the NTT and WHT data subsets. Because HD 29697 is an active star (Casares et al. 1996), this might hint that the companion star in Cen X-4 is also active. Fig.~\ref{fig2} depicts the normalized Doppler-shifted average of Cen X-4 before and after subtraction of the K5~V template. The Cen X-4 spectrum shows H$\alpha$ and \hbox{He\,{\sc i}}~$\lambda6678$ emission lines. \hbox{Li\,{\sc i}}~$\lambda6708$ in absorption becomes prominent after the subtraction. The detection of a strong \hbox{Li\,{\sc i}}~$\lambda6708$ line in the late-type companion is an unexpected result since the initial \hbox{Li\,{\sc i}} content should be quickly depleted by means of convective mixing in the star and mass transfer to the compact object. The reader should refer to Mart\'\i n et al. (1994) for a detailed discussion of the significance of the anomalous high abundance of \hbox{Li\,{\sc i}} in Cen X-4 and other XRNe.   

For the rotational broadening calculation we have selected a subset of our data which has the higher spectral resolution. This consists of the target and template spectra acquired with the WHT. The spectra were Doppler-shifted to the rest frame of the template HR 6159 and then the target spectra were summed with different weights to optimize the final signal-to-noise ratio. In order to correctly estimate $v \sin i$ we simulate in our template spectra the drift due to the orbital motion of the secondary during the exposures: we created 8 versions of each WHT template which were smeared in the same way as the target spectra by convolution with a rectangular profile of the appropriate width. This is given by:

\[
\sigma_{\rm smear}=t_{\rm exp}{2 \pi\over P} {K_2} | \cos {\left({2 \pi}\varphi \right) |}
\]
where $t_{\rm exp}$ is the length of the exposure and $\varphi$ the orbital phase.

The smeared template copies were then summed using identical weights as for the Cen X-4 spectra. Subsequently, we broadened the three summed templates from 1 to 100 km s$^{-1}$, in steps of 1 km s$^{-1}$, and subtracted the resultant broadened spectra (as above) from the target Doppler-corrected average. The summed templates were broadened through convolution with the rotational profile of Gray (1992). We used linearized limb-darkening coefficients of $\mu=0.69$, $\mu=0.73$ and $\mu=0.75$ appropriate for a K3 III, K5 III and K7 III star, respectively (Al-Naimiy 1978; Wade \& Rucinski 1985). The results from the $\chi^2_\nu$ minimization are listed in Table~\ref{tablevsini}, with quoted uncertainties on $v \sin i$ corresponding to $\chi^{2}_{\rm min}$+1. Because the minimum $\chi^2_\nu$ for the output residuals is acceptable ($\simeq1$) for the three templates, we choose for the rotational broadening the mean of the results  obtained from the analysis: $v \sin i = 43 \pm 6$ km s$^{-1}$. To check the systematic errors introduced by the choice of $\mu$ (which is only suitable for the continuum; Collins \& Truax 1995), we have allowed $\mu$ to vary in the range 0-1. This leads to 5 per cent ($\sim$ 3 km s$^{-1}$) changes in the resulting value of $v \sin i$, and therefore it is the statistical noise and not the uncertainty in $\mu$ which limits our accuracy. Unfortunately, we cannot make use of the higher signal-to-noise spectra acquired with the AAT and NTT to refine the rotational broadening. The absorption features of the former are artificially broadened by the instrumental resolution (74 km s$^{-1}$), whereas the latter are dominated by the instrumental resolution (45 km s$^{-1}$) and the orbital smearing. In particular, the total broadening (calculated by adding in quadrature the instrumental resolution and $\sigma_{\rm smear}$) ranges from 25 to 37~km s$^{-1}$ for the WHT spectra and 47 to 61~km s$^{-1}$ for the NTT spectra.

\section{THE BINARY SYSTEM PARAMETERS}

Assuming that the companion star is synchronized with the binary motion and fills its Roche lobe we can calculate the mass ratio ($q = M_{2}/M_{1}$) through the expression (see e.g. Wade \& Horne 1988):
\[ 
v \sin i = 0.462~K_{2}~q^{1/3} (1 + q)^{2/3}
\]
From the values of $v \sin i$ and $K_2$, we derived $q=0.17 \pm 0.06$, which implies a primary velocity semi-amplitude of ${K_1}=q{K_2} = 26 \pm 9$ km s$^{-1}$. On the other hand, our values of $K_2=150.0\pm1.1$ km s$^{-1}$ and $P=0.6290496\pm0.0000021$~d imply a mass function of $f(M)=0.220\pm0.005$~M$_\odot$. Combining the values for $f(M)$ and $q$, the masses of the compact object and the companion star are: $M_1=(0.30\pm0.04)~\sin^{-3}i$~M$_\odot$ and $M_2=(0.05 \pm 0.03)~\sin^{-3}i$~M$_\odot$. We used the constraint upon the inclination from the IR observations of Shahbaz et al. (1993). By fitting the {\it H}-band light curve with an ellipsoidal model these authors found the system inclination to be in the range 31$^{\circ} < i < 54^{\circ}$ (90 per cent confidence). Taking the extreme limits for $i$ we obtain ranges for the masses of: $0.49 < M_1 < 2.49$ M$_\odot$ and $0.04 < M_2 < 0.58$ M$_\odot$. The companion star of Cen X-4 is of spectral type ${\rm K}3-{\rm K}5$ V. Main-sequence stars of these spectral types have masses in the range $0.74-0.68$~M$_\odot$. The disagreement with the masses obtained above is easily explained if the secondary star is undermassive (i.e. evolved) for its spectral type as proposed by Shahbaz et al. (1993).  

\section{DOPPLER TOMOGRAPHY}

\begin{figure}
\epsfig{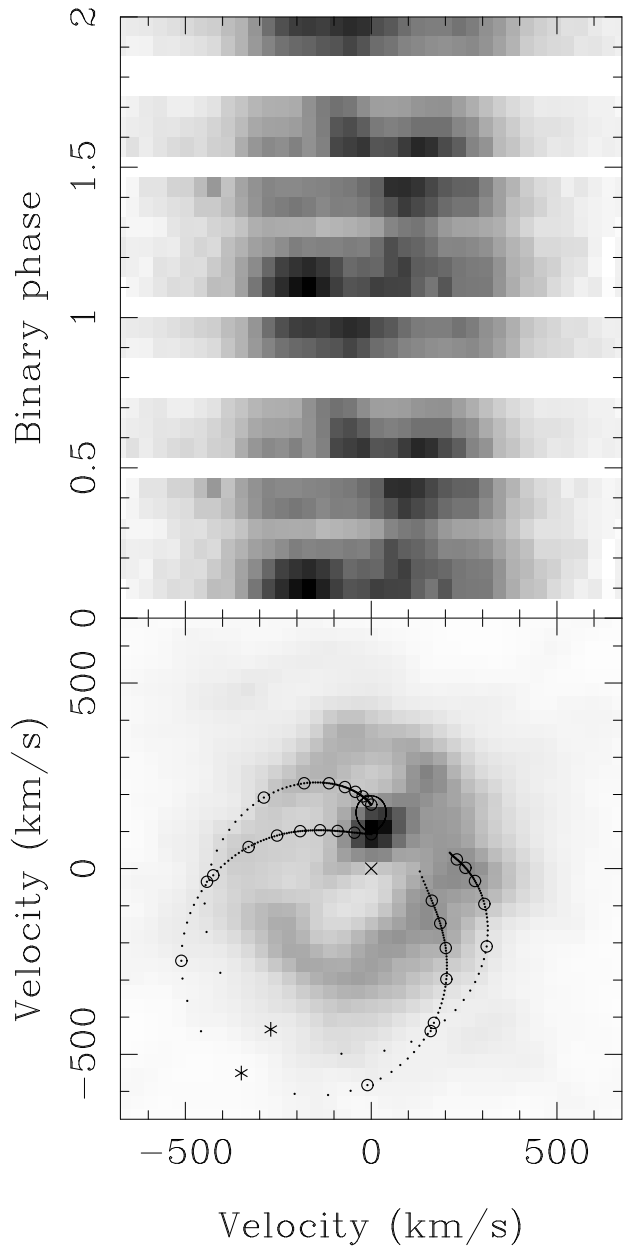}
\caption{Doppler image of the H$\alpha$ emission (bottom panel). The data, represented in the form of a trailed spectra, are depicted in the top panel. Empty strips represent gaps in the phase coverage. For the sake of clarity, the same cycle has been plotted twice. The Roche lobe of the secondary star, the predicted velocities of the gas stream (lower curve) and of the disc along the stream (upper curve) are plotted for $K_2=150$ km s$^{-1}$ and $q=0.17$. Distances in multiples of 0.1$R_{\rm L1}$ are marked along both curves with open circles. The centre of mass of the system is denoted by a cross. Note that the Doppler image has been computed from the original data without subtracting the spectrum of the secondary star, improper subtraction of which can lead to an erroneous indication of emission from the secondary (see e.g. Marsh et al. 1994).}
\label{fig3}
\end{figure}

The H$\alpha$ emission line has a FWHM of $640$ km s$^{-1}$ and a mean
equivalent width (EW) of 35 $\pm$ 7 \AA, with the error accounting for
the intrinsic line variability during the three runs. The line profile
has a double-peaked structure, the classic signature of an accretion
disc, and shows periodic changes in the height of the two peaks by the
action of an S-wave emission component. We have made use of the
Doppler tomography technique (Marsh \& Horne 1988) on the orbitally
resolved line profiles to reconstruct the H$\alpha$ brightness
distribution and identify the location of the S-wave in the binary
system. In order to construct the Doppler map, the 25 spectra were
re-binned into 82 pixels of constant velocity scale (36.6 km s$^{-1}$)
and the continua removed. The derived Doppler image computed with the
maximum entropy method (Marsh \& Horne 1988) is shown in the bottom
panel of Fig.~\ref{fig3}. This image should be interpreted as a one
year time-averaged map since the data were acquired over a year. We
have overplotted the theoretical path of the gas stream and the Roche
lobe of the secondary star and the gas stream trajectory for $K_2=150$
km s$^{-1}$ and $q=0.17$. The Doppler image shows a region of enhanced
intensity at the position of the secondary star, superposed on a weak
ring-like structure, which represents the emission arising from a
rotating accretion disc. There is no evidence for emission from the
gas stream or the hotspot (the stream/disc impact region). We think
unlikely that this absence is an artifact of using spectra spread over
a year since a very similar result is obtained when using only the AAT
data subset which expands over 5 consecutive orbital
cycles. Individual Doppler maps for the WHT and NTT data subsets could
not be computed because of scarce phase coverage. However, we note
that the S-wave component from the secondary is clearly detected in
all the individual line profiles.

\section{DISCUSSION}

\begin{figure}
\epsfig{width=6in,file=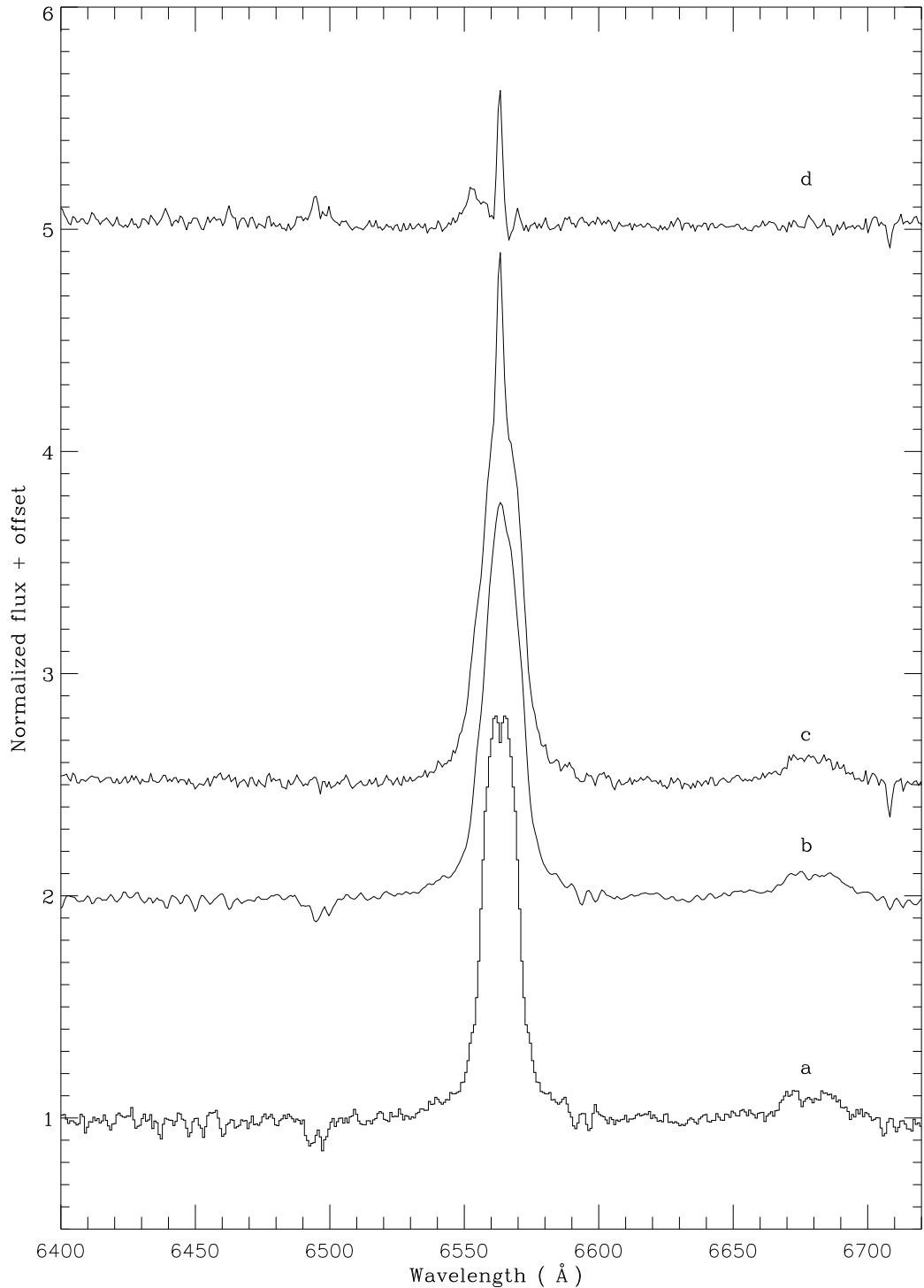}
\caption{Detrending the narrow H$\alpha$ component. The simulated
symmetric H$\alpha$ profile (a) has been smeared according to the
orbital phases of the individual spectra (b). Removal of (b) from the
Doppler-corrected average spectrum of Cen X-4 in the rest frame of the
secondary after subtraction of the ${\rm K}5$ V template (c) yields
the narrow H$\alpha$ component (d). Arbitrary vertical offsets have
been applied  to the spectra for the sake of clarity.}
\label{fig4}
\end{figure}

In order to constrain the possible causes of the H$\alpha$ emission arising from the secondary, we have extracted the H$\alpha$ S-wave component from the line profiles as described by Casares et al. (1997). The procedure applied is illustrated in Fig.~\ref{fig4}. First, we simulate the accretion disc by averaging all the spectra in the laboratory rest frame. Because the resultant profile showed the blue peak significantly stronger than the red one (clearly an artifact of the S-wave emission due to the gaps in our phase coverage), we built a symmetric double-peaked profile by folding a copy of the red part of the emission line profile (spectrum $a$ in Fig.~\ref{fig4}). Next, twenty five versions of this 'accretion disc' profile were shifted to the rest frame of the secondary star, at the times of our individual spectra, and then averaged using the same weights (spectrum $b$ in Fig.~\ref{fig4}). Finally, the resultant profile was removed from the Doppler-corrected average of Cen X-4 with the ${\rm K}5$ V template spectrum subtracted, to be left with the narrow H$\alpha$ component with EW(H$\alpha$)=1.7$\pm$0.5~\AA~(spectrum $d$ in Fig.~\ref{fig4}). In an attempt to find secular variations in the strength of the H$\alpha$ S-wave component, we extracted it (as above) from each individual data subset. It was found that the values of EW(H$\alpha$) were comparable within the errors to the one obtained using the complete baseline.

To compare EW(H$\alpha$) with observations of rapidly rotating stars, we convert this value into H$\alpha$ line surface flux at the star ({\it F}$_{{\rm H}\alpha}$). {\it F}$_{{\rm H}\alpha}$ is derived using the relationship given by Soderblom et al. (1993): 
\begin{eqnarray*}
\log F_{{\rm H}\alpha} & = & \log {\rm EW}({\rm H}\alpha) + 0.113(B-V)^{2}_{0} \\
                       &   & - 1.188(B-V)_{0} + 7.487
\end{eqnarray*}
Using ${\it (B-V)}_{0}$=+1.25 mag (Chevalier et al. 1989) and an
equivalent width (corrected by the disc's contribution) of
EW(H$\alpha$)$=1.7/0.75=2.3$~\AA, we obtain {\it F}$_{{\rm
H}\alpha}=3.5\times10^{6}$ erg cm$^{-2}$ s$^{-1}$. The dependence of
the chromospheric activity on rotation is given by the relation
between the fraction of the stellar luminosity emitted in the
H$\alpha$ line ($R_{{\rm H}\alpha} = F_{{\rm H}\alpha} / \sigma
T^{4}_{\rm eff}$) and the Rossby number $R_{0}=P/ \tau_{\rm c}$ (Noyes
et al. 1984), where $P$ is the orbital period of the star and
$\tau_{\rm c}$ is the convective turnover time. Adopting $T_{\rm
eff}=3600$~K (Shahbaz et al. 1993) and $\tau_{\rm c}\simeq 25$~d, as
calculated using eq. 4 of Noyes et al. (1984), we derive $\log R_{{\rm
H}\alpha} = -3.4$ and $\log R_{0} = -1.6$. These values can be
compared with observations of rapidly rotating ($P\geq0.25$~d)
chromospherically active stars in the Pleiades (Soderblom et al. 1993;
fig. 19a). According to fig.~19a, saturation of chromospheric activity
would occur for $P\leq$~2d or $v \sin i \geq 15$ km s$^{-1}$. Indeed
the companion of Cen X-4, with $P=0.63$~d and  $v \sin i = 43$ km
s$^{-1}$, lies in the region of activity saturation ($\log R_{{\rm
H}\alpha} \simeq -3.7$), but with $R_{{\rm H}\alpha}$ slightly higher
than expected.

We now examine the possibility that the H$\alpha$ component is caused by the incidence of X-rays on the inner face of the secondary. The observations of Cen X-4 have shown that its quiescent X-ray luminosity is variable on time-scales from days (Campana et al. 1997) to years (van Paradijs et al. 1987; Rutledge et al. 2001). We adopt the 1994 intrinsic X-ray luminosity of $2.4\times10^{32}$ erg s$^{-1}$ ($0.5-10$~keV) obtained with {\it ASCA} (Asai et al. 1996). Using this value and neglecting any shielding by the accretion disc, the X-ray flux irradiation of the secondary star would be $F_{\rm X}=3\times10^{8}$ erg cm$^{-2}$ s$^{-1}$, where we have used a binary separation of 3.6 R$_{\odot}$ (assuming ${M_1}=1.4$~M$_\odot$). At these flux levels, the secondary would be receiving sufficient energy to power the narrow H$\alpha$ component. Therefore we can only conclude that the H$\alpha$ emission arising from the secondary star may be either due to chromospheric activity, and/or powered by X-ray heating on the secondary.

H$\alpha$ emission with origin in the secondary star has also been found in the quiescent systems Nova Muscae 1991 ($P=10.4$~hr, $v \sin i=106$ km s$^{-1}$; Casares et al. 1997) and probably in A0620-00 ($P=7.8$~hr, $v \sin i=83$ km s$^{-1}$; Marsh et al. 1994), V404 Cygni ($P=155.3$~hr, $v \sin i=39$ km s$^{-1}$; Casares \& Charles 1994; Casares 1996), Nova Ophiuchi 1977 ($P=12.5$~hr, $v \sin i \le 79$ km s$^{-1}$; Harlaftis et al. 1997) and Nova Scorpii 1994 ($P=62.9$~hr, $v \sin i= 86$ km s$^{-1}$; Shahbaz et al. 1999). The high rotational broadening measured suggests that coronal activity may exist in these XRNe. Unfortunately, the poor phase coverage in the case of Nova Sco 1994 and Nova Oph 1977, the lack of EW values of the H$\alpha$ component in A0620-00 and V404 Cyg, and the fact that we only have an upper limit to the X-ray flux of Nova Mus 1991, prevent us from doing a more detailed analysis of the origin of the H$\alpha$ emission in these systems.

The contribution to the optical flux due to a hotspot is required in order to explain and model the optical and IR light curves of Cen X-4 (McClintock \& Remillard 1990; Shahbaz et al. 1993). The presence of such hotspot was strongly supported by the phasing of the 1987 H$\beta$ emission line radial velocity curve (McClintock \& Remillard 1990). Therefore its absence in our H$\alpha$ tomogram is a puzzling result. A possible explanation would be that the hotspot is fainter or comparable in brightness to the accretion disc at H$\alpha$ and brighter at shorter wavelengths (higher members of the Balmer series or high excitation lines). A second possibility is that the secondary star in Cen X-4 might under-fill its Roche lobe by a few percent of its radius and mass transfer is unstable.

The acquisition of new orbitally resolved high-resolution data will bring substantial progress in the knowledge of Cen X-4. A simultaneous and uniform orbital phase coverage of the binary orbit may reveal whether or not the origin of the secondary star's H$\alpha$ emission is only chromospheric or due to X-ray heating. The absence of hotspot at H$\alpha$ in XRNe in quiescence has also been seen in Nova Persei 1992 ($P=5.09$~hr; Harlaftis et al. 1999), Nova Mus 1991 (Casares et al. 1997) and maybe in Nova Oph 1977 (Harlaftis et al. 1997). This result suggests that it may be a common phenomena among X-ray transients. Unfortunately very little data following an XRN during its quiescent state are presently available. Therefore observations at different epochs of quiescence are required to show whether this is a transient phenomenon or a temperature effect of the hotspot.

\section{SUMMARY}

We have presented spectroscopic observations of the XRN Cen X-4 during quiescence. A $\chi^2$ test applied to the residuals obtained by subtracting different template stars from the Cen X-4 spectrum supports a secondary star of spectral type ${\rm K}3-{\rm K}5$ V contributing $\simeq75$~per cent of the observed flux at H$\alpha$. Using the high-resolution spectra, we have determined the rotational broadening of the companion star to be $v \sin i = 43 \pm 6$ km s$^{-1}$. This has provided a reliable  measurement of the mass ratio in Cen X-4: $q=0.17 \pm 0.06$. We obtain a mass function of $0.220\pm0.005$~M$_\odot$ which, when combined with the mass ratio and binary inclination constraints, gives a range of $0.49 < M_1 < 2.49$ M$_\odot$ and $0.04 < M_2 < 0.58$ M$_\odot$ for the neutron star and secondary star masses. This result reinforces the possibility that the secondary is evolved for its spectral type. Furthermore, we have revised the ephemeris of Cen X-4 for the inferior conjunction of the secondary to ${T_0}=\rm{HJD}~2449163.934(1)+0.6290496(21)\times N$~d. Our Doppler tomogram of the H$\alpha$ line is dominated by a strong emission arising from the companion star but no hotspot contribution is observed. The H$\alpha$ flux derived from the S-wave component is consistent with the value expected for a chromospherically active and/or X-ray heated secondary.

\section*{ACKNOWLEDGMENTS}

We are grateful to P. Molaro and E. Mart\'\i{}n for helping during the NTT observations. We also thank the referee, Dr Daniel Rolfe, for useful comments. Use of {\sc molly}, {\sc doppler} and {\sc trailer} routines developed largely by T. R. Marsh is acknowledged. This research has made use of the SIMBAD database, operated at CDS, Strasbourg, France.

\clearpage
\begin{table}
\caption{Journal of Observations.}
\label{tablelog}
\begin{center}
\begin{tabular}{lccccc}
Date & Telescope & No. spectra & Exp. time & $\lambda$ range & Dispersion \\ 
& & & (s) &  (\AA) & (\AA~pix$^{-1}$) \\
\\
24-27 Jun 1993 &    AAT	  &	   11	      &       1800	   &   6155-6968     &        0.80	        \\
6 Mar 1994 	&    NTT          &	    3	      &       2400	   &   6175-6815     &         0.31	        \\
7 Mar 1994 	&     "           &	    2	      &     1800-1900	   &       "         &  	 "	        \\
8 Mar 1994 	&     "           &	    1	      &       3300	   & 	   "	     &  	 "	        \\
9 Jul 1994 	&    WHT          &	    2	      &       1800	   &   6360-6767     &       
 0.40	        \\
10 Jul 1994 	&     "           &	    2	      &     1200-1800	   &        "        &  	 "	        \\     
13 Jul 1994 	&     "           &	    2	      &       1800	   &        "        &  	 "	        \\
14 Jul 1994   &     "           &	    2	      &         "	   &   6307-6714     &  	 "              \\ 
\end{tabular}
\end{center}
\end{table}

\begin{table}
\caption{Radial velocity parameters (errors at 1$\sigma$).}
\label{tablerv}
\begin{center}
\begin{tabular}{ccccccc}
Template &Spectral &$\gamma$      &$\gamma_{_2}^{a}$         &$K_{_2}$         &$T_0$                           & $\chi^2_{\nu}$\\
         &Type     &(km s$^{-1}$) &(km s$^{-1}$)      &(km s$^{-1}$)    & (+2249163)                   & (d.o.f=22)     \\
\\
HR 5536  & K3~III   &207.4$\pm$0.9 & 187.0 $\pm$ 2.9   &152.2 $\pm$ 1.0  & 0.934 $\pm$ 1$\times$10$^{-3}$ & 7.1  \\
HR 6159  & K4~III   &180.9$\pm$1.0 & 184.0 $\pm$ 1.9   &150.0 $\pm$ 1.1  & 0.934 $\pm$ 1$\times$10$^{-3}$ & 4.5  \\
HR 5622  & K5~III   &200.5$\pm$0.8 & 185.4 $\pm$ 1.7   &151.0 $\pm$ 1.0  & 0.934 $\pm$ 1$\times$10$^{-3}$ & 6.8 \\
\end{tabular}
\begin{tabular}{c}
\\
$^a$ $\gamma_{_2}$: Systemic velocity after adding the radial velocity of the standard star found in {\sc SIMBAD}.
\end{tabular}
\end{center}
\end{table}

\begin{table}
\caption{Spectral classification.}
\label{tablespeclas}
\begin{center}
\begin{tabular}{lccccc}
Template     &Spectral     &$\chi^2_{\nu}$   &$f$       \\
 	     &Type	   &(d.o.f.=159)     &           \\ 
\\
HR 5536         &    K3~III	& 1.92  	      & 0.72 $\pm$ 0.03\\ 
HR 6159         &    K4~III	& 1.97                & 0.76 $\pm$ 0.03\\
HR 5622 	&    K5~III	& 2.11	              & 0.64 $\pm$ 0.02\\	
HD 184467       &     K2~V	& 1.75                & 1.00 $\pm$ 0.03 \\  
HD 29697        &     K3~V	& 1.30                & 0.72 $\pm$ 0.02\\       
HD 154712 A     &     K4~V	& 1.48                & 0.76 $\pm$ 0.03\\
61 Cyg A        &     K5~V	& 1.34                & 0.75 $\pm$ 0.03\\
61 Cyg B        &     K7~V	& 1.58                & 0.69 $\pm$ 0.02\\ 
\end{tabular}
\end{center}
\end{table}     

\begin{table}
\caption{Rotational broadening.}
\label{tablevsini}
\begin{center}
\begin{tabular}{cccccc}
Template  &Spectral	 &$\mu$       &$v \sin i$      &$\chi^2_{{\rm min},\nu}$      \\
          &Type          &            &(km s$^{-1}$)   &        (d.o.f.=321)     \\
\\
HR 5622           &    K3~III	 &  0.69      &42.5 $\pm$ 6.0& 0.97  \\
HR 6159           &    K4~III	 &  0.73      &44.2 $\pm$ 5.4&0.97   \\
HR 5536           &    K5~III	 &  0.77      &42.0 $\pm$ 5.8&1.00   \\
\end{tabular}
\end{center}
\end{table}

\clearpage

\clearpage

\clearpage

\clearpage


\begin{thebibliography}{} 

\bibitem[Al-Naimiy(1978)]{aln78}Al-Naimiy H. M., 1978, Ap\&SS, 53, 181 

\bibitem[Asai et al.(1996)]{asa96} Asai K., Dotani T., Mitsuda K., Hoshi R., Vaughan B., Tanaka Y., Inoue H., 1996, PASJ, 48, 257
	
\bibitem[Campana et al.(1997)]{cam97} Campana S., Mereghetti S., Stella L., Colpi M., 1997, A\&A, 324, 941

\bibitem[Canizares, McClintock \& Gridlady(1980)]{can80} Canizares C. R., McClintock J. E., Gridlady J. E., 1980, ApJ, 236, L55

\bibitem[Casares \& Charles(1994)]{cas94}Casares J., Charles P. A, 1994, MNRAS, 271, L5
	
\bibitem[Casares et al.(1996)]{cas96a}Casares J., Mouchet M., Mart\'\i nez-Pais I. G., Harlaftis E. T., 1996, MNRAS, 282, 182

\bibitem[Casares(1996)]{cas96b}Casares J., 1996, in Evans A.,  Wood J. H., eds,  in IAU Colloq. 158, Cataclysmic Variables and Related Objects. Kluwer, Dordrecht, p. 395
	        
\bibitem[Casares et al.(1997)]{cas97}Casares J., Mart\'\i{}n E. L., Charles P. A., Molaro P., Rebolo, R., 1997, NewA, 1, 299


\bibitem[Charles(1998)]{cha98}Charles P. A., 1998, in Abramowicz M., Bjornsson G., Pringle J., eds, Theory of Black Hole Accretion Disks. Cambridge Univ. Press, Cambridge, p.1

\bibitem[Chevalier et al.(1989)]{che89}Chevalier C., Ilovaisky S. A., van Paradijs J., Pedersen H., van der Klis M., 1989, A\&A, 210, 114 

\bibitem[Collins \& Truax(1995)]{col95}Collins G. W. II., Truax R. J., 1995, ApJ, 439, 860
        
\bibitem[Conner, Evans \& Belian(1969)]{con69}Conner J. P., Evans W. D., Belian R. D., 1969, ApJ, 157, L157

\bibitem[Cowley et al.(1988)]{cow88}Cowley A. P., Hutchings J. B., Schmidtke P. C., Hartwick F. D. A., Crampton D., Thompson I. B., 1988, AJ, 95, 1231
 	
\bibitem[Gray(1992)]{gra92}Gray D. F., 1992, The Observation and Analysis of Stellar Photospheres. Cambridge University Press, Cambridge

\bibitem[Harlaftis et al.(1997)]{har97}Harlaftis E., Steeghs D., Horne K., Filippenko A. V., 1997, AJ, 114, 1170

\bibitem[Harlaftis et al.(1999)]{har99}Harlaftis E., Collier S., Horne K., Filippenko A. V., 1999, A\&Ap, 341, 491
        	
\bibitem[Lampton, Margon \& Bowyer(1976)]{lam76}Lampton M., Margon B., Bowyer S., 1976, ApJ, 208, 177

\bibitem[McClintock \& Remillard(1990)]{mcc90}McClintock J. E., Remillard R. A., 1990, ApJ, 350, 386        
        	
\bibitem[Marsh \& Horne(1988)]{mar88}Marsh T. R., Horne K., 1988, MNRAS, 235, 269
        		
\bibitem[Marsh, Robinson \& Wood(1994)]{mar94}Marsh T. R., Robinson E. L., Wood J. H., 1994, MNRAS, 266, 137
        		
\bibitem[Mart\'\i n et al.(1994)]{mart94}Mart\'\i n E. L., Rebolo R., Casares J., Charles P. A., 1994, ApJ, 435, 791	
        		
\bibitem[Matsuoka et al.(1980)]{mat80}Matsuoka M. et al., 1980, ApJ, 240, L137	
        	
\bibitem[Noyes et al.(1984)]{noy84}Noyes R. W., Hartmann L. W., Baliunas S. L., Duncan D. K., Vaughan A.H., 1984, ApJ, 279, 763

\bibitem[Rutledge et al.(2001)]{rut01}Rutledge R. E., Bildsten L., Brown E. F., Pavlov G. G., Zavlin V. E., 2001, ApJ, 551, 921	

\bibitem[Shahbaz, Naylor \& Charles(1993)]{sha93}Shahbaz T., Naylor T., Charles P. A., 1993, MNRAS, 265, 655	

\bibitem[Shahbaz et al.(1999)]{sha99}Shahbaz T., van der Hooft F., Casares J., Charles P. A., van Paradijs, 1999, MNRAS, 306, 89

\bibitem[Soderblom et al.(1993)]{sod93}Soderblom D. R., Stauffer J. R., Hudon J. D., Jones B. F., 1993, ApJS, 85, 315  

\bibitem[Tanaka \& Shibazaki(1996)]{tan96}Tanaka Y., Shibazaki N., 1996, ARA\&A, 34, 607

\bibitem[van Paradijs et al.(1987)]{van87}van Paradijs J. Verbunt F., Shafer R. A., Arnaud K. A., 1987, 182, 47 
l
\bibitem[van Paradijs \& McClintock(1995)]{van95}van Paradijs J., McClintock, J. E. 1995, in Lewin W. H. G., van Paradijs J. \& van den Heuvel E. P. J., eds, X-Ray Binaries. Cambridge University Press, Cambridge, p. 58
 
\bibitem[van Paradijs(1998)]{van98}van Paradijs J., 1998, Buccheri R., van Paradijs J., Alpar M. A., eds, The Many Faces of Neutron Stars. Kluwer, Dordrecht, p. 279  
	
\bibitem[Wade \& Rucinski(1985)]{wad85}Wade R. A., Rucinski S. M., 1985, AApS, 60, 471
	
\bibitem[Wade \& Horne(1988)]{wad88}Wade R. A., Horne K., 1988, ApJ, 324, 411
	
\end{thebibliography}
\end{document}